\documentclass[twocolumn,showpacs,preprintnumbers,amsmath,amssymb]{revtex4}

\usepackage{graphicx}

\def\r{\rho}
\def\be#1{\begin{equation}  \label{#1}}          
\def\ee{\end{equation}}

\def\Gl#1{{Eq.~(\ref{#1})}}

\def\Fig#1{{Fig.~\ref{#1}}}

\begin{document}

\title{Morphological Thermodynamics of Fluids: Shape Dependence of Free Energies}

\author{P.-M. K\"onig}
\author{R. Roth}
\author{K. R. Mecke}

\affiliation{Max-Planck-Institut f\"ur Metallforschung, Heisenbergstr. 3,
  70569 Stuttgart, Germany} 
\affiliation{Institut f\"ur Theoretische und Angewandte Physik,
  Pfaffenwaldring 57, 70550 Stuttgart, Germany}

\begin{abstract}
We examine the dependence of a thermodynamic potential of a fluid on 
the geometry of its container. If motion invariance, continuity,
and additivity of the potential are fulfilled, only four morphometric 
measures are needed to describe fully the influence of an arbitrarily shaped 
container on the fluid. These three constraints can be understood as
a more precise definition for the conventional term "extensive" and
have as a consequence that the surface tension and other thermodynamic 
quantities contain, beside a constant term, only contributions 
linear in the mean and Gaussian
curvature of the container and not an infinite number of curvatures as 
generally assumed before. We verify this numerically in the
entropic system of hard spheres bounded by a curved wall.
\end{abstract}

\maketitle

Although thermodynamics is built on extremely general
assumptions its implications are far reaching and
powerful. One basic building block is geometry which has
a long history in thermodynamics and statistical physics
of condensed matter. The formulation of thermodynamics in
terms of differential forms \cite{frankel}, scaled-particle theory
(SPT) for fluids \cite{Reiss60}, depletion forces \cite{Roth00} of colloids in
biological cells \cite{Ellis01}, and density functional theory \cite{Evans79} 
(DFT) based on fundamental geometric measures \cite{Rosenfeld89,Roth02,Yu02} 
are only a few examples for the importance of a general geometric point
of view on thermodynamic properties. 
In Refs. \cite{likos95,mecke96a,mecke96b} the structure and phase 
behavior of microemulsions was explained assuming that the free energy of the 
fluid on a mesoscopic scale is given solely by four fundamental geometric 
measures \cite{mecke94}. Here, we show numerically for the first time that
these four fundamental measures are sufficient to describe accurately
the free energy of a hard-sphere fluid in contact with a complexly shaped wall
which supports the more general assumptions made in 
Refs.~\cite{likos95,mecke96a,mecke96b,mecke94}.

The grand potential $\Omega=\Omega[S; T,\mu]$ of a fluid depends on the
temperature $T$ and the chemical potential $\mu$ of the system, as
well as on certain geometrical quantities which describe
the shape of the container that bounds the system $S$. 
What are these thermodynamically relevant
morphological parameters? One usually argues that every
thermodynamic potential is an extensive quantity, which
means that it scales linearly with the "size" of the
system $S$. By partitioning a large system into identical
smaller subsystems one normally assumes that $\Omega[S; T,\mu]$ is
proportional to the volume $V=V[S]$ of the system and uses as
ansatz $\Omega[S; T, \mu] = \omega(T,\mu) \cdot V[S]$ . 
The intensive quantity $\omega(T,\mu)$ is the negative of the pressure $p(T,\mu)$ 
and is independent of the "size" of the confining container of $S$. 
This simple ansatz however is only valid for
infinite bulk, i.e. ``border-less'' systems.
If $S$ is bounded by a container, $\Omega$ depends
on the shape of the container in a potentially
complicated manner and is conventionally described by an infinite expansion
in powers of the curvatures of the wall.  However, we show that general
considerations restrict this functional dependence on 
the shape to a linear combination of only four
morphological measures, if all intrinsic (correlation) length scales
are small compared to the system "size". This finding is
in particular important for systems such as porous media \cite{arnsnature},
biological cells \cite{Ellis01}, or complex fluids such as microemulsions 
\cite{likos95} where fluids are confined by
complexly shaped compartments and where the dependence of
thermodynamic quantities and transport properties on the
shape of pores or cells has significant functional and
biological consequences. Our arguments can however not be
applied, to critical phenomena, or if long
ranged fluid-fluid or fluid-wall interactions are
considered or if wetting or drying phenomena \cite{Evans03} occur at
the wall, as intrinsic lengths in such systems have a
macroscopic size. 

Now we focus on the dependence of $\Omega[S]$ on the shape of the
system $S$. The actual form of this mapping from a container onto a real number is
given by the type and state of the fluid under
consideration and is a complicated integral over the
phase-space of the system, which usually can only be
calculated approximately. However, we impose the
following three physical restrictions on this mapping:

{\bf (i) Motion\ invariance:} 
The thermodynamic potential of a system must be
independent of its location and orientation in space,
i.e. $\Omega[g S]=\Omega[S]$ for all translations and 
rotations $g$ in $3$ dimensions.

{\bf (ii) Continuity:} 
If a sequence of convex sets $S_n$ converges towards the convex set 
$S$ for $n\rightarrow \infty$, then $\Omega[S_n]\rightarrow \Omega[S]$.
Intuitively, this continuity property expresses the fact that
an approximation of a convex domain by e.g. convex polyhedra
also yields an approximation of the thermodynamic potential $\Omega[S]$ by 
$\Omega[S_n]$. In SPT it is 
shown that continuity is violated if the container $S$ is very small
and comparable in size to that of fluid particles \cite{reiss59}. Here we
consider a larger container $S$.

{\bf (iii) Additivity:} The functional of the union $S_1\cup S_2$
of two domains $S_i$, $i=1,2$, is the sum of the
functional of the single domains subtracted by the  
intersection: $\Omega[S_1 \cup S_2] = \Omega[S_1] + \Omega[S_2] - \Omega[S_1 \cap S_2]$.
This relation generalizes the common rule for the addition of an extensive 
quantity for two disjunct domains $S_1 \cap S_2 = \emptyset$ to the case of overlapping domains by 
subtracting the value of the thermodynamic quantity of the double-counted 
intersection. Note that the intersection $S_1 \cap S_2$ does not need to be a
volume but can rather be an area or a line for adjacent containers
$S_i$. Additivity can break down if long ranged interactions are present 
or if the system develops a macroscopic
intrinsic length scale. A fluid can be considered additive even inside a
concave container, if opposing walls are separated by several correlation 
lengths.

Naturally the question arises about the most general form
of a potential that satisfies these three conditions. The
Hadwiger theorem \cite{hadw1,ijmb98} states that every motion-invariant,
continuous and additive functional in $3$ dimensions
can be written as a linear combination of the volume $V=\int_{S} dV$,
the surface area $A=\int_{\partial S} dA$, the integrated mean curvature $C=\int_{\partial S} H dA$, and
the Euler characteristic $X=\int_{\partial S} K dA$ of the
container. Therefore we write 
\be{eq:MorphometricOmega}
\Omega[S] = -p V[S] + \sigma A[S] + \kappa C[S] + \bar \kappa X[S]
\ee
as a {\it complete} expression for the grand canonical
potential, if the aforementioned conditions are satisfied 
\cite{likos95,mecke96a,mecke96b,mecke94}. The pressure 
$p(T,\mu)$, the surface tension at the planar wall
$\sigma(T,\mu)$, and the bending rigidities $\kappa(T,\mu)$ and $\bar
\kappa(T,\mu)$ are properties of the fluid and the wall-fluid interaction, but
are independent of the actual shape of the bounding wall. 
The latter two thermodynamic coefficients describe the 
influence of the curvature of the wall. 
Similar coefficients are also used for the Helfrich Hamiltonian 
\cite{helfrich73}, which describes the free energy cost of bending a 
membrane. It is compatible with \Gl{eq:MorphometricOmega} on
a length scale larger than the persistence length of the membrane, where 
renormalized contributions proportional to $H^2$ vanish due to 
thermal fluctuations.

Note that \Gl{eq:MorphometricOmega} can be easily applied even to 
complexly shaped objects because the
shape of $S$ enters $\Omega$ only via the four simple
morphometric measures $V$, $A$, $C$, and $X$, while the thermodynamic 
coefficients $\sigma$, $\kappa$ and $\bar \kappa$ can be determined in a simple
geometry.

Thermodynamic quantities can be derived directly from $\Omega$ 
and inherit a simple dependence on the shape of $S$ by
virtue of \Gl{eq:MorphometricOmega}. The interfacial tension 
$\gamma=(\Omega + p V)/A$, which measures the total change in the grand 
potential per unit area introduced by the wall, can be evaluated using 
\begin{equation} \label{eq:gamma}
\gamma = \sigma + \kappa \bar H + \bar \kappa \bar K,
\end{equation}
where $\bar H = C/A$ and $\bar K = X/A$ are the
averaged mean and Gaussian curvatures of the bounding
wall. These geometrical quantities can be calculated from
the principal radii of curvature $R_1$ and $R_2$ via 
$H=(1/R_1 + 1/R_2)/2$ and $K=1/(R_1 R_2)$. Note that
this further justifies the ansatz used in SPT \cite{Reiss60,Bryk03} for the
interfacial tension and shows that the analytic
dependence of $\gamma$ on the curvature is
a direct consequence of the additivity of the grand
potential. No higher powers or
derivatives of $\bar H$  or $\bar K$ contributing either 
to $\gamma$ or to $\Omega$. 

\begin{figure}
\centering
\includegraphics{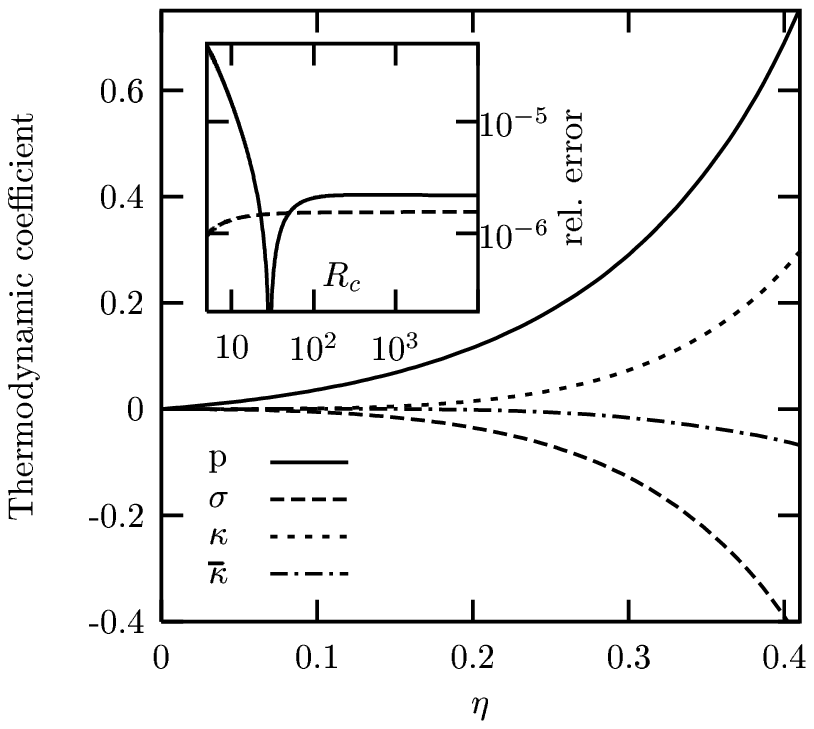}
\caption{\label{fig:thermoqu} 
The expansion coefficients as defined in \Gl{eq:MorphometricOmega} of the
grand potential of a hard sphere fluid. For each value of $\eta$, the four
thermodynamic coefficients can be used to calculate thermodynamic quantities
for arbitrarily shaped systems. The solid line of the inset shows the relative
error of $\gamma$ at a cylinder with radius $R_c$
calculated using the thermodynamic coefficients at $\eta=0.3$. This error can
be compared to the numerical relative error for a sum rule
\cite{Henderson83,Bryk03} 
(dashed line), which gives an estimate for the accuracy of our DFT data. Both
errors are of the same order of magnitude such that our numerical data is in
agreement with the prediction of \Gl{eq:MorphometricOmega}.  
The fluid was modeled via Rosenfeld's FMT \cite{Rosenfeld89}.}
\end{figure}

Closely related to the interfacial tension is $\Gamma$, the excess (over the bulk)
amount of fluid adsorbed at the wall per unit area, which is defined as  
$\Gamma \equiv (1/A)\int_{S} (\rho({\bf r}) - \rho)\;dV$, 
where $\rho$ is the bulk density and $\rho({\bf r})$ is the inhomogeneous
density distribution of the fluid. It inherits the morphometric form
of the interfacial tension via Gibbs' adsorption theorem 
\begin{equation}
-\Gamma=\left(\frac{\partial \gamma}{\partial \mu}
\right)_{T,V} = \frac {\partial \sigma}{\partial \mu} + \frac {\partial \kappa} {\partial \mu} \bar H +
\frac {\partial \bar \kappa}{\partial \mu} \bar K. 
\end{equation}

If a fluid is bounded by a container one denotes the number density
closest to the wall as contact density $\r^c$, which becomes $\bar \r^c$,  
when averaged over the boundary surface. For hard walls $\bar \r^c$ can be
regarded as a thermodynamic quantity because of an exact sum-rule
\cite{Henderson83,Bryk03} and also features a simple
dependence on the geometry: 
\begin{equation} \label{eq:rhoc}
\bar \r^c = p + 2 \sigma \bar H + \kappa \bar K.
\end{equation}
This relation can be derived by generalizing the arguments
given by Henderson \cite{Henderson83}.
Note that the density distribution of the fluid away from contact depends in a
more complicated way on the curvature of the wall.

\begin{figure}
\centering
\includegraphics{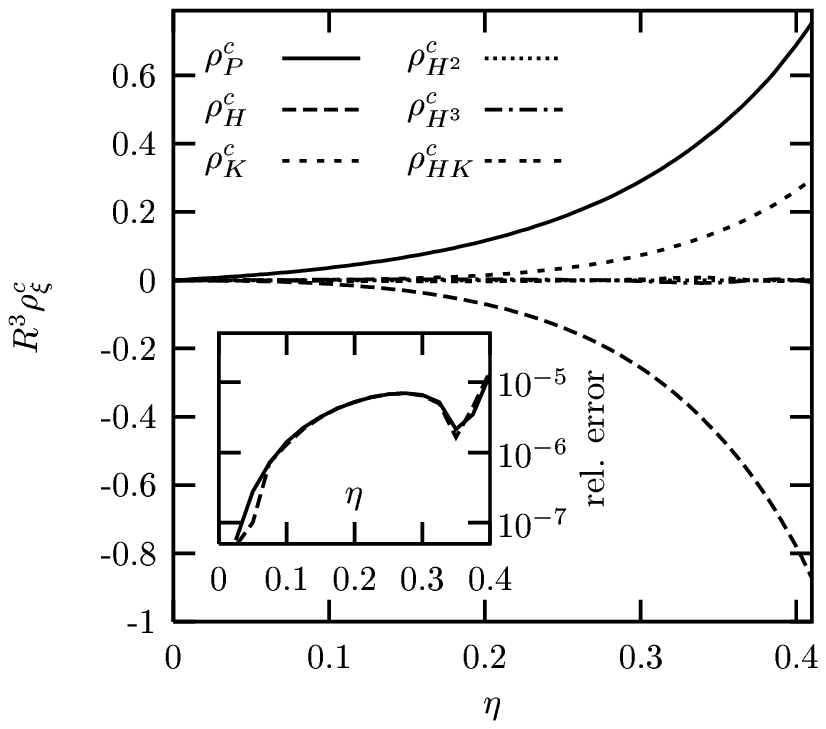}
\caption{\label{fig:contactexp}
Curvature expansion coefficients for the contact density of a hard-sphere
fluid as a function of  the packing fraction $\eta$. Only the additive 
contributions $\rho^c_P$, $\rho^c_H$, and $\rho^c_K$ contribute. The inset 
shows the relative error for the contact density at a spherical container with
radius $5 R$ between a direct DFT calculation and two approaches: For the 
solid curve all shown curvature expansion coefficients have been used, for the
dashed one only additive contributions. The error, which is due to the 
numerical inaccuracies, is in both cases roughly the same. This shows that 
non-additive terms do not contribute to the thermodynamic quantity $\rho^c$. 
Similar results were found for $\gamma$ (see \Fig{fig:thermoqu}).} 
\end{figure}

We test the vanishing of higher powers of $\bar H$ and $\bar K$ in
thermodynamic quantities such as $\Omega$, $\gamma$, $\Gamma$, or $\bar \r^c$ 
by extensive numerical studies of a fluid of hard spheres of radius $R$ bounded
by a hard wall using DFT. 
For a given chemical potential $\mu$, the bulk number density $\r$ of
spheres or equivalently the 
packing fraction $\eta=4 \pi R^3 \r /3$ are fixed. Hard-core interactions do
not introduce an energy scale such that we can account
for the temperature dependence of all
thermodynamic quantities by scaling energies with $\beta=1/(k_B T)$.
One can expect the grand potential of a hard-sphere fluid to be additive as 
hard core interactions are short ranged and the correlation length remains 
small in the fluid phase. Unfortunately, for arbitrarily shaped
walls, there exists no reliable and accurate direct
method to calculate thermodynamic quantities. However,
for simple geometries such as planar, spherical, or
cylindrical walls, it is possible to apply e.g. DFT
techniques \cite{Evans79}, which allow the 
calculation of $\sigma$, $\kappa$ and $\bar \kappa$. 
For these geometries, the curvatures and the
contact density are constants over the surface such that
we can replace averages over the surface by their local
quantities. Rosenfeld's fundamental measure theory (FMT) \cite{Rosenfeld89}
and the White-Bear version of FMT \cite{Roth02,Yu02} have proven to
describe the thermodynamics and structure of
hard-sphere fluids very accurately and both
give qualitatively equivalent results in the following
analysis, as well as the non FMT-based Tarazona Mark I functional
\cite{tmark1}. It is important to realize that minimizing a FMT functional
does not restrict the results to standard SPT behavior \cite{Bryk03}.

The thermodynamic coefficients of \Gl{eq:MorphometricOmega} are
shown in \Fig{fig:thermoqu} as functions of the packing fraction $\eta$ of
the fluid. As dividing interface, which determines the surface
area $A$ as well as the curvatures $H$ and $K$, we choose
the surface where the contact density is measured.
From these coefficients one can obtain values for $\gamma$ and
$\rho^c$ for various geometries. 
In the inset of
\Fig{fig:thermoqu} we show the relative error (full line) of 
$\gamma$ of a hard-sphere fluid with packing fraction
$\eta=0.3$ at a cylinder with radius $R_c$ as calculated with the thermodynamic
coefficients compared to that obtained directly from DFT. This relative error
is of the same order of magnitude as the relative numerical error of the 
contact sum-rules, Eq.~(\ref{eq:rhoc}), indicating clearly that the very small
deviation between the morphometric interfacial tension and that from a direct
DFT calculation is a numerical error of our calculation.

In order to show that the linear expansion in $H$ and $K$ of 
$\gamma$ and $\rho^c$ is sufficient, we
include for the analysis of our numerical data also non-additive terms and
show that they do not contribute either to $\gamma$ or to $\rho^c$. To this
end we introduce the curvature expansion of e.g. the contact density 
\begin{eqnarray} \label{eq:cecontact}
\r^c & = &\r^c_P + \r^c_H H + \r^c_K K + \r^c_{H^2} H^2 + 
\r^c_{HK} HK \\
& +& \r^c_{H^3} H^3 + {\cal O}(R_{1,2}^{-4}). \nonumber
\end{eqnarray}
The coefficients $\r^c_\xi$ for $\xi= P, H, K, H^2, HK, H^3, \dots$ can be 
determined numerically using DFT. The results in \Fig{fig:contactexp} show that
irrespective of the bulk density only the additive terms $\rho^c_P$,
$\rho^c_H$, and $\rho^c_K$ contribute to $\r^c$. Note that in particular
$\r^c_K$ (additive) contributes to the contact density whereas $\r^c_{H^2}$
(non-additive) does not although both $K$ and $H^2$ feature the same quadratic
dependence on the radii of curvature. This shows that the expansion in $H$
and $K$ is complete and not a truncated power series. In the inset of
\Fig{fig:contactexp} we show the relative error of the contact density of a
hard-sphere fluid at a sphere with radius $5 R$ as a function of $\eta$
obtained by the morphometric form (dashed line) of $\rho^c$, 
Eq.~(\ref{eq:rhoc}), and by the generalized curvature expansion (full line), 
Eq.~(\ref{eq:cecontact}), both compared to the direct calculation of
DFT. Since in both cases the relative error is roughly the same 
we conclude that additive contributions are sufficient to describe the
influence of curvature on the contact density $\rho^c$.

To further highlight that the remarkably simple functional dependence on only
four morphometric measures of the container is a peculiar feature
of thermodynamic quantities, we introduce, in addition to the hard-wall
potential, a short-ranged {\em soft} part to the wall-fluid potential of the
form  $V_{soft}(u)=V_0 \exp(-u/\lambda)$, where $u$ denotes the normal distance
from the dividing interface. It can be either attractive, if
$V_0<0$, or repulsive, if $V_0>0$. Note that this additional part of the
wall-fluid interaction leaves the morphometric form of thermodynamic
quantities such as the grand potential $\Omega$,
Eq.~(\ref{eq:MorphometricOmega}), and the surface tension $\gamma$,
Eq.~(\ref{eq:gamma}), unchanged. The contact density $\rho^c$ however loses its
morphometric form for any non-zero value of $V_0$ because the sum-rule
Eq.~(\ref{eq:rhoc}) acquires an additional term \cite{Henderson83}.

\begin{figure}
\includegraphics{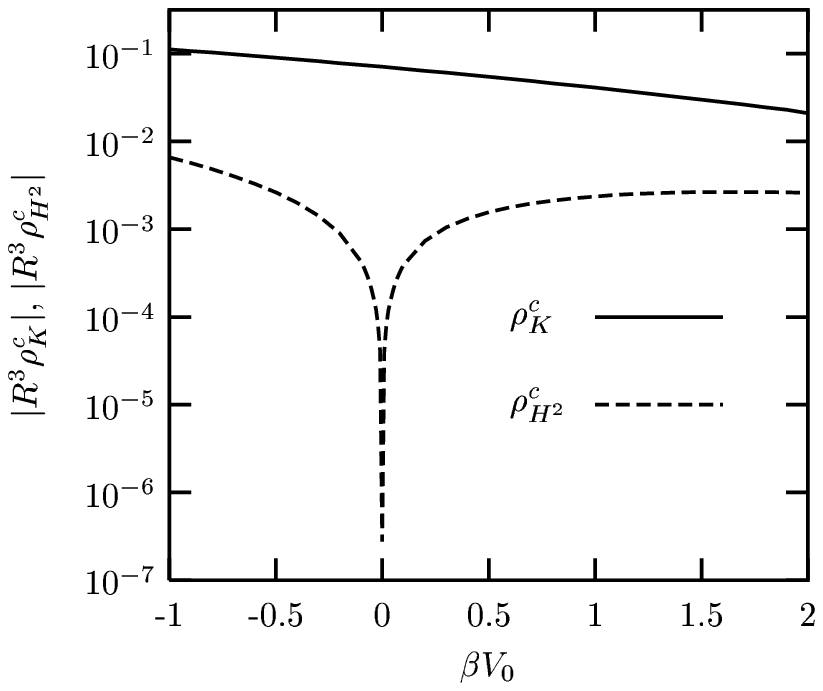}
\caption{\label{fig:extrenalpot}
If the external potential of a hard container is perturbed by an additional 
potential as e.g. $V_{soft}(u)=V_0 \exp(-u/\lambda)$ the contact density
$\r^c$ is no longer a thermodynamic quantity for $V_0 \neq 0$. In this case
non-additive contributions such as $\r^c_{H^2}$ are necessary to 
describe the dependence on the curvatures of the wall. 
Here we show $\r^c_K$ and $\r^c_{H^2}$ in a
logarithmic plot for $\lambda=R$ and a packing fraction $\eta=0.3$. It is
striking that even very small perturbations lead to $\r^c_{H^2} \neq 0$ 
which is a clear indication that additivity of the contact density $\r^c$
is destroyed. In contrast, $\r^c_K$ contributes to the contact density for all $V_0$.}
\end{figure}

This effect can be demonstrated by performing a curvature expansion,
Eq.~(\ref{eq:cecontact}), of the contact density as a function of $V_0$. In the 
case of a purely hard wall, $V_0 = 0$, 
the curvature expansion contains only additive contributions
($\r^c_P$, $\r^c_H$, and $\r^c_K$), whereas for non-zero values of $V_0$ this is no longer 
the case and the amplitude of e.g. $\r^c_{H^2}$, increases with an increasing value of
$|V_0|$. As an example of this behavior we display in Fig.~\ref{fig:extrenalpot}
$\rho^c_K$ (additive) and $\rho^c_{H^2}$ (non-additive) for a hard-sphere fluid with packing fraction
$\eta=0.3$ for values of $\beta V_0$ from $-1$ to $2$ and
$\lambda=R$. It is very striking that already for very small amplitudes $|V_0|
\neq 0$, for which the soft part of the wall-fluid interaction can be
regarded as a small perturbation, the contact density $\rho^c$ acquires
additional non-additive contributions. We verified that the
interfacial tension $\gamma$ keeps its morphometric form (not shown), as
expected. 
 
Monte-Carlo simulation confirm that the contact density 
of a hard-sphere fluid with  $\eta=0.3314$
around a biaxial ellipsoid with half axes $(4,4,10) R$ as obtained  
by \Gl{eq:rhoc}. Within the
statistical errors of the simulation (about $1\%$), the results of \Gl{eq:rhoc}
together with the thermodynamic coefficients shown in \Fig{fig:thermoqu} agree
with the simulation data \cite{bryk04}.

Furthermore, we also confirmed that \Gl{eq:gamma} holds accurately in concave
geometries if the perturbations introduced by the container do 
not interfere at caustic points. For a relatively 
high packing fraction of hard spheres of $\eta=0.4$ the deviation between
\Gl{eq:gamma} and direct DFT results is smaller than $1.5\%$ 
for all radii of the cylinder larger than $5 R$.

We presented a more precise definition than the
conventional term "extensive" to describe the dependence
of a thermodynamic quantity of a fluid with short
intrinsic length scales on the shape of the system. The
assumption that the grand potential of a fluid is
motion-invariant, continuous and additive allows an
expansion in terms of only four simple morphological
functionals \cite{likos95,mecke96a,mecke96b,mecke94}. As a consequence, the 
curvature expansion for thermodynamic quantities terminates after
linear terms in mean and Gaussian curvature. This
observation allows a calculation of thermodynamic
quantities for complexly shaped objects with a greatly
reduced effort in comparison to direct methods. The ideas
presented here and tested numerically for the hard-sphere fluid 
should be applicable also to fluids with short-ranged interactions provided
that intrinsic length scales are small compared to typical
features of the container. If however, length scales are comparable, continuity
or additivity may be violated and then thermodynamic quantities 
acquire additional contributions.

The present approach is currently being extended to include local structure
(density profiles) which will allow the prediction of entropic contributions
to the effective interactions of complicated macromolecules.

We thank H. Wagner, R. Evans and J.R. Henderson for stimulating discussions.

\bibliography{paper}

\newpage
\end{document}